\documentclass[aps,prd,onecolumn,groupedaddress,showpacs,nofootinbib,amssymb]{revtex4}
\usepackage{graphicx,bm}
\usepackage[english]{babel}
\usepackage{amsmath}
\usepackage{amssymb}
\usepackage{amsfonts}

\begin{document}

\def\cL{{\cal L}}
\def\be{\begin{equation}}
\def\ee{\end{equation}}
\def\bea{\begin{eqnarray}}
\def\eea{\end{eqnarray}}
\def\beq{\begin{eqnarray}}
\def\eeq{\end{eqnarray}}
\def\tr{{\rm tr}\, }
\def\nn{\nonumber \\}
\def\e{{\rm e}}

\title{Could dynamical Lorentz symmetry breaking induce the superluminal
neutrinos?}

\author{Shin'ichi Nojiri$^{1,2}$
and Sergei D. Odintsov$^3$\footnote{Also at Tomsk State Pedagogical
University, Tomsk, Russia}}

\affiliation{
$^1$ Department of Physics, Nagoya University, Nagoya
464-8602, Japan \\
$^2$ Kobayashi-Maskawa Institute for the Origin of Particles and
the Universe, Nagoya University, Nagoya 464-8602, Japan \\
$^3$Instituci\`{o} Catalana de Recerca i Estudis Avan\c{c}ats
(ICREA) and Institut de Ciencies de l'Espai (IEEC-CSIC), Campus
UAB, Facultat de Ciencies, Torre C5-Par-2a pl, E-08193 Bellaterra
(Barcelona), Spain}

\begin{abstract}

A toy fermion model coupled to the  Lagrange multiplier constraint field
is proposed. The
possibility of superluminal neutrino propagation as a result of dynamical
Lorentz symmetry breaking is studied.

\end{abstract}


\maketitle

The OPERA experiment results indicate towards the possibility that the
neutrino speed might
exceed the speed of light \cite{:2011zb}. Although the experimental results 
could
be in principle fully trustable,
they have not been yet confirmed  by other experiments. Hence, the superluminal
neutrino occurrence remains at present as a quite hypothetical case,
especially in  relation with the emerging relativistic causality violation.
However, it might be interesting to consider an eventual realization of such
a theoretical effect in quantum field theory. Several attempts, mainly  related 
to the Lorentz symmetry breaking to give some theoretical explanation for
superluminal neutrino, have just appeared \cite{speculations}. 
For for earlier proposals, see \cite{Pas:2005rb}.

In the present note we propose a toy model of fermion theory coupled to the 
Lagrange multiplier constrained  field, which induces the dynamical Lorentz
symmetry breaking.
As a result, the superluminal neutrino propagation appears to be possible.
The description of the model is motivated by the analogy with
the power-counting renormalizable gravity proposed in
Ref.~\cite{Horava:2009uw}.
In this model, the Lorentz symmetry or full diffeomorphism invariance is
explicitly broken and the dispersion relation of the
graviton is modified so that the UV behavior of the quantum field could be
improved.
Due to the modified dispersion relation, the speed of the graviton can be
faster than the light speed.

However, the lack of the full diffeomorphism invariance leads to the extra
scalar mode in
the model \cite{Horava:2009uw}. Due to the presence of scalar mode,
the general relativity and/or the Newton law cannot be reproduced even in the
IR region.

In order to solve this problem, the models with the full diffeomorphism
invariance have
been investigated \cite{Nojiri:2009th,Kluson:2011rs}. The diffeomorphism
invariance and/or Lorentz symmetry can be spontaneously broken since the
derivative of the scalar field with respect to the coordinates does not vanish.
The non-vanishing value is generated by the Lagrange multiplier field, which
gives a constraint on the derivative of the scalar field not to vanish.
The mechanism is very similar to the St\"uckelberg
formulation of the massive U(1) gauge theory. For the model 
\cite{Kluson:2011rs},
it has been explicitly shown that the UV behavior of graviton is improved as in
the original theory \cite{Horava:2009uw} but due to the full diffeomorphism
invariance, there does not appear the extra (scalar/vector) mode.
In such covariant gravity \cite{Kluson:2011rs}, the other fields besides the
gravity sector are assumed
to be standard ones, what guarantees the experimentally observed Lorentz
invariance.
Moreover, in such a theory \cite{Kluson:2011rs}, the Lorentz symmetry in the
matter sector could be
broken if we consider the intermediate states including graviton though the
breakdown could be very small.

Let us show, however, that we can construct a model of spinor which might be
identified with neutrino, where the speed of the spinor can exceed the light 
speed.
In such a construction, the Lorentz symmetry is broken spontaneously as in the
models \cite{Nojiri:2009th,Kluson:2011rs}. We will estimate the parameters to
be roughly consistent with the OPERA experiment data\cite{:2011zb}.

We now start with the action including the Lagrange multiplier field
$\lambda$ \cite{vikman} and the scalar field $\phi$:
\be
\label{Lag}
S_\mathrm{Lag} = - \int d^4 x \lambda \left( \frac{1}{2}
\partial_\mu \phi \partial^\mu \phi
+ U_0 \right) \, ,
\ee
which gives a constraint
\be
\label{LagHL2}
\frac{1}{2} \partial_\mu \phi \partial^\mu \phi
+ U_0 = 0\, ,
\ee
that is, the vector $(\partial_\mu \phi)$ is time-like.
Therefore, the Lorentz symmetry is broken spontaneously.
This mechanism was used to construct a power-counting renormalizable and
covariant model of gravity \cite{Nojiri:2009th,Kluson:2011rs}.
One may choose the direction of time to be parallel to the vector
$(\partial_\mu \phi)$ and
in the following, we assume
\be
\label{ppert1}
\phi = \sqrt{2 U_0} t\, .
\ee
Here the gravity sector is not included. We would like to  construct a model
of spinor,
whose speed can exceed the light speed, although the action has the full
Lorentz symmetry.
The action we consider is
\be
\label{lvf1}
S = \int d^4 x \left[
\bar\psi \left\{ \gamma^\mu \partial_\mu
+ \alpha \left( P_\mu^{\ \nu} \gamma^\mu \partial_\nu\right)^{2n+1} \right\}
\psi - \lambda \left( \frac{1}{2}
\partial_\mu \phi \partial^\mu \phi
+ U_0 \right)\right] \, .
\ee
Here $\alpha$ is a constant, $n$ is an integer equal to or greater than $1$,
and $P_\mu^{\ \nu}$ is a projection operator defined by \cite{Kluson:2011rs}
\be
\label{ppert2}
P_\mu^{\ \nu} \equiv \delta_\mu^{\ \nu} + \frac{\partial_\mu \phi
\partial^\nu \phi}{2U_0}\, .
\ee
The equation corresponding to the Dirac equation has the following form:
\be
\label{lvf2}
0=\left\{ \gamma^\mu \partial_\mu
+ \alpha \left( P_\mu^{\ \nu} \gamma^\mu \partial_\nu\right)^{2n+1} \right\} \psi\, .
\ee
By using (\ref{ppert1}), Eq.~(\ref{lvf2}) looks as:
\be
\label{lvf3}
0=\left\{ \gamma^0 \partial_0 + \gamma^i \partial_i
+ \alpha \left( \gamma^i \partial_i \right)^{2n+1} \right\} \psi\, .
\ee
The dispersion relation for the spinor is then given by
\be
\label{lvf4}
\omega = k \sqrt{ 1 + \alpha^2 k^{4n} }\, .
\ee
Here $\omega$ is the angular frequency corresponding to
the energy and $k$ is the wave number corresponding to the momentum.
In the high energy region, the dispersion relation becomes
\be
\label{lvf5}
\omega \sim |\alpha| k^{2n+1} \, ,
\ee
and therefore the phase velocity $v_p$ and the group velocity $v_g$ are given, 
respectively, by
\be
\label{lvf6}
v_p \equiv \frac{\omega}{k} = |\alpha| k^{2n}\, ,\quad
v_g \equiv \frac{d\omega}{dk} = \left(2n+1\right) |\alpha| k^{2n}\, .
\ee
When $k$ becomes larger, both  $v_p$ and $v_g$ become also larger
in an unbounded way and exceed the light speed.

When the breakdown of the Lorentz symmetry is small, we may expand (\ref{lvf4})
as follows:
\be
\label{lvf7}
\omega \sim k \left( 1 + \frac{\alpha^2 k^{4n}}{2} \right)\, .
\ee
Since we choose the light speed $c$ to be unity, $c=1$,
the OPERA experiment \cite{:2011zb} shows
\be
\label{lvf8}
\frac{v-c}{c} = \frac{\alpha^2 k^{4n}}{2}
= (2.48 \pm 0.28\, (\mbox{stat.})\, \pm 0.30\, (\mbox{sys.}))
\times 10^{-5}\, ,
\ee
for
\be
\label{lvf9}
k \sim E_\nu = 17\, \mbox{GeV}\, .
\ee
Here $v$ and $E_\nu$ are the speed and the energy of the neutrino,
respectively.
Then one finds
\be
\label{lvf10}
\alpha^{- \frac{1}{2n}} \sim 10^{1 + \frac{5}{4n}}\, \mbox{GeV}\, .
\ee
This suggests that the scale of the Lorentz symmetry breaking could be
$10-100\, \mbox{GeV}$.
By using (\ref{lvf8}) and (\ref{lvf9}), we may rewrite the deviation from
the light speed as
\be
\label{lvf11}
\frac{v-c}{c} = \frac{\alpha^2 k^{4n}}{2}
= 2.48 \times 10^{-5} \left( \frac{k}{17\, \mbox{GeV}}\right)^{4n}
\, .
\ee
When $k\sim 10\, \mbox{MeV}$, a stringent limit was given by the observation of
(anti-)neutrinos emitted by the SN1987A supernova \cite{Hirata:1987hu}:
\be
\label{lvf12}
\frac{|v-c|}{c} < 2 \times 10^{-9}\, ,
\ee
which could be consistent due to the $k$-dependence in (\ref{lvf11}).
For $k\sim 10\, \mbox{MeV}$, Eq.~(\ref{lvf11}) gives
\be
\label{lvf13}
\frac{v-c}{c} \sim 10^{-5-16n}\, .
\ee
Then the constraint Eq.~(\ref{lvf12}) can be easily satisfied even for $n=1$.

If the spinor (\ref{lvf1}) is identified with neutrino, the usual derivatives 
should
be replaced
by the covariant derivatives which include gauge bosons and therefore new kinds 
of
interactions would emerge. Such a interaction could be suppressed by the
$k$-dependence
for low energy region. In the high energy region, however, the corrections 
could be
large. This completes the construction of our toy model for superluminal
neutrino. Of course, a number of questions remain to be understood if such a 
model
has any  relation with reality. For instance, the interpretation of Lagrange
multiplier scalars introduced originally as the dust of dark energy 
\cite{vikman}
should be developed. Also why only neutrino seems to show a superluminal 
behavior? Moreover, effects related to the manifestation of superluminal
propagation at very high energies should be searched for.

Recently there was a claim that suplerluminal neutrino would lose energy 
rapidly due to the bremsstrahlung of electron-positron pairs \cite{Cohen:2011hx}. 
For the model without Lorentz invariance, the high energy neutrino can decay 
into the low energy neutrino itself and other particles like electron-positron 
pairs, which is not yet strictly proven. 
Such a process is prohibited for the Lorentz invariant models since we can 
always choose the coordinate frame that (massive) neutrino is not moving, 
where the decay to the neutrino itself and other particle violates 
the energy conservation.
In our model, since there is a Lorentz invariance in the Lagrangian although 
the invariance is spontaneously broken, there exists local conserved energy 
momentum tensor $T_{\mu\nu}$. The conservation law has the standard Lorentz 
invariant form $\partial^\mu T_{\mu\nu}=0$. Then the existence of the 
Lorentz invariantly conserved energy momentum tensor might prohibit the process 
that the high energy neutrino could decay into the low energy neutrino itself 
and other particles although we need more detailed analysis. 

Moreover, in Ref.~\cite{Li:2011rt}, it was proposed the different scenario 
of background dependent violation of the Lorentz invariance. The proposed scenario is 
consistent with OPERA experiment and it passes the bremsstrahlung problem
\cite{Cohen:2011hx}. The model under discussion in this work may be further 
generalized to the one with spontaneous background dependent Lorentz symmetry violation.
This will be considered elsewhere.

\section*{Acknowledgments \label{VI}}

We are much  grateful to M.~Chaichian for sharing our interests and for very
helpful discussions.
This research has been supported in part
by MEC (Spain) project FIS2006-02842 and AGAUR(Catalonia) 2009SGR-994 (SDO),
by Global COE Program of Nagoya University (G07)
provided by the Ministry of Education, Culture, Sports, Science \&
Technology and by the JSPS Grant-in-Aid for Scientific Research (S) \# 22224003
and (C) \# 23540296 (SN).

\end{document}